\documentclass[12pt]{article}
\usepackage{amsmath,amsfonts,amssymb,graphicx}
\usepackage{graphics, setspace}

\usepackage{epstopdf}
\DeclareGraphicsExtensions{.eps}

\begin{document}
\thispagestyle{empty}

\def\theequation{\arabic{section}.\arabic{equation}}
\def\a{\alpha}
\def\b{\beta}
\def\g{\gamma}
\def\d{\delta}
\def\dd{\rm d}
\def\e{\epsilon}
\def\ve{\varepsilon}
\def\z{\zeta}
\def\B{\mbox{\bf B}}\def\cp{\mathbb {CP}^3}

\newcommand{\h}{\hspace{0.5cm}}

\begin{titlepage}

\renewcommand{\thefootnote}{\fnsymbol{footnote}}
\begin{center}
{\Large \bf Bethe states for the two-site Bose-Hubbard model: a binomial approach}
\end{center}
\vskip 1.2cm \centerline{\bf Gilberto Santos$^{1}$, Changrim  Ahn$^{2}$, Angela Foerster$^{3}$ and Itzhak Roditi$^{1}$}

\vskip 10mm
\centerline{\sl$\ ^1$ Centro Brasileiro de Pesquisas F\'{\i}sicas - CBPF}
\centerline{\sl Rua Dr. Xavier Sigaud, 150, Urca, Rio de Janeiro - RJ - Brazil}
\vskip .5cm
\centerline{\sl$\ ^2$Department of Physics} \centerline{\sl Ewha Womans
University} \centerline{\sl DaeHyun 11-1, Seoul 120-750, S. Korea}
\vskip .5cm
\centerline{\sl$\ ^3$Instituto de F\'{\i}sica}
\centerline{\sl Universidade Federal do Rio Grande do Sul}
\centerline{\sl Av. Bento Gon¸calves, 9500, Agronomia, Porto Alegre - RS - Brazil}

\vspace*{0.6cm} \centerline{\tt gfilho@cbpf.br, ahn@ewha.ac.kr,}
 \centerline{\tt angela@if.ufrgs.br and roditi@cbpf.br}
\vskip 20mm

\baselineskip 18pt

\begin{center}
{\bf Abstract}
\end{center}

We calculate explicitly the Bethe vectors states by the algebraic Bethe ansatz method with the  $gl(2)$-invariant $R$-matrix for the two-site Bose-Hubbard model. Using a binomial expansion of the $n$-th power of a sum of two operators we get and solve a recursion equation. We calculate the scalar product and the norm of the Bethe vectors states. The form factors of the imbalance current operator are also  computed.

\end{titlepage}
\newpage
\baselineskip 18pt

\def\nn{\nonumber}
\def\tr{{\rm tr}\,}
\def\p{\partial}
\newcommand{\non}{\nonumber}
\newcommand{\bea}{\begin{eqnarray}}
\newcommand{\eea}{\end{eqnarray}}
\newcommand{\bde}{{\bf e}}
\renewcommand{\thefootnote}{\fnsymbol{footnote}}
\newcommand{\be}{\begin{eqnarray}}
\newcommand{\ee}{\end{eqnarray}}

\vskip 0cm

\renewcommand{\thefootnote}{\arabic{footnote}}
\setcounter{footnote}{0}

\setcounter{equation}{0}
\section{Introduction}

The first experimental verification of the Bose-Einstein condensation (BEC) \cite{ak,anderson,wwcch} occurred after a gap of more than seven decades following its theoretical prediction \cite{bose,eins}. After its realization a great deal of progress has taken place both in the theoretical and experimental study of this physical phenomenon \cite{Dalfovo,l01,Bagnato,cw,Donley,Piza,Bloch,Carusotto}. A particularly fruitful instrument in relation to ultracold physics are many-body atomic models related to BEC. In this direction the quantum inverse scattering method (QISM) \cite{fst,ks1,takhtajan,korepin,faddeev} has been used to solve and study some prototypical many-body models that  contribute to describe phenomena associated to BEC \cite{GSantos11a,GSantos11b,GSantos13}.  Some of these models, despite their simplicity, display a rich structure showing quantum phase transitions and interesting semi-classical behaviour that have been studied in \cite{GSantos06a,GSantos09,GSantos10, Duncan,Zhou1,Zhou2}, and explored in different areas such as nuclear physics, condensed matter and atomic-molecular physics. To keep things as simple as possible we shall consider here the two-site Bose-Hubbard, also known in special cases as the canonical Josephson Hamiltonian \cite{l01}. This model may be viewed as a particular case of the bosonic multi-state two-well model studied in \cite{GSantos13}, and can be used to describe a  quadrupolar nuclei system in nuclear magnetic resonance  \cite{NMR} by a $N/2$ Schwinger pseudo-spin realization of the Hamiltonian. Conversely, there is also a link with the Schwinger bosonic realization of the Lipkin-Meshkov-Glick model \cite{Lipkin,Dukelsky} used to study closed shells in a nuclei model. In spite, of course, from being a two-site specialization of the Bose-Hubbard model it is a very useful model in various realms as the understanding of tunnelling phenomena using two BEC \cite{albiez,milb,hines2,ours,our,hines,GSantos06b}, as well as quantum phase transitions  using tools of quantum computation and quantum information.  The model is described by the Hamiltonian 
\begin{equation}
\hat{H} = \frac{K}{8}(\hat{N}_1 - \hat{N}_2)^2 - \frac{\Delta \mu}{2}(\hat{N}_1 -\hat{N}_2)
 - \frac{\mathcal{E}_J}{2}(\hat{a}_1^\dagger \hat{a}_2 + \hat{a}_2^\dagger \hat{a}_1),
\label{ham} 
\end{equation}
\noindent where, $\hat{a}_1^\dagger, \hat{a}_2^\dagger$, denote the single-particle creation
 operators in each site and,  $\hat{N}_1 = \hat{a}_1^\dagger \hat{a}_1,
 \hat{N}_2 = \hat{a}_2^\dagger \hat{a}_2$, are the corresponding
 boson number operators. The total boson particles number operator, $\hat{N} = \hat{N}_1+\hat{N}_2$,
 is a conserved quantity, $[\hat{H},\hat{N}]=0$. The coupling $K$ provides the interaction strength between the 
 bosons and is proportional to the $s$-wave scattering length,  $\Delta \mu$ is the external potential and ${\cal E}_J$ is the amplitude of tunnelling.
 
The Hamiltonian (\ref{ham}) is integrable in the sense that it can be solved by the quantum inverse scattering method (QISM) and it has been discussed in different ways using this method  \cite{GSantos06b,jlletter,jlreview,Angela, AngelaEric,jlsigma,jlbroot,Angela2,ATonel}. The algebraic formulation of the Bethe ansatz, associated to the QISM, was primarily  developed in  \cite{fst,ks1,takhtajan,korepin,faddeev}.

A very important problem in the algebraic Bethe ansatz method is the construction of the Bethe vectors states (BVS) \cite{jlletter,ks2,slavnov} using the correspondent creation operator applied to the pseudo-vacuum. Employing this form of the BVS it is possible \cite{slavnov} to calculate their scalar product and then use it to calculate important physical quantities as the form factors. Form factors are defined as the matrix entries of operators in the base of the eigenvectors of the Hamiltonian. Another important application is in the calculation of the average values of the operators as for example correlation operators. Applying this method, some physical quantities for the Hamiltonian (\ref{ham}) were obtained in \cite{jlletter}. Recently, BVS have shown to be useful in fundamental issues of planar ${\cal{N}} = 4$ super Yang-Mills (SYM) theory  \cite{Gromov1,Gromov2} in the context of the integrability in the AdS/CFT correspondence \cite{niklas-ahn}.

In the present work, we develop and use a new method to explicitly calculate the BVS  and obtain the scalar product of two BVS for the two-site Bose-Hubbard model. Although we concentrate on this model the procedure is of general applicability. We use a Lax operator to construct a realization of the monodromy matrix and get an algebraic identity between  the associated $C$-operator and the $D$-operator of the monodromy matrix (see next section) needed to calculate the BVS. We then use the binomial expansion for the $n$-th power of the sum of two operators: in a first step we will show that the binomial expansion can be written as a sum of permutations of the product of that operators or as a standard binomial expansion as in a commutative algebra plus a function of the commutator of these two operators; in a second step we write a recursion equation and give its solution. Next, we calculate the scalar product between one on-shell and one generic off-shell BVS, as well as the norm. As an application we obtain the form factors (non normalized) for the imbalance current operator.

\newpage
\section{The algebraic Bethe ansatz method}

The spectrum of the Hamiltonian (\ref{ham}) has appeared in different papers \cite{GSantos06b,jlletter,jlreview,Angela,jlsigma,jlbroot,Angela2,ATonel} using this method.  To fix notation we will shortly describe the algebraic Bethe ansatz method, see \cite{jlreview,Angela,Roditi} for more details.  We begin with the $gl(2)$-invariant $R$-matrix, depending on the spectral parameter $u$,

\begin{equation}
R(u)= \left( \begin{array}{cccc}
1 & 0 & 0 & 0\\
0 & b(u) & c(u) & 0\\
0 & c(u) & b(u) & 0\\
0 & 0 & 0 & 1\end{array}\right),
\end{equation}

\noindent with $b(u)=u/(u+\eta)$, $c(u)=\eta/(u+\eta)$ and $b(u) + c(u) = 1$. Above,
$\eta$ is an arbitrary parameter, to be chosen later.

It is easy to check that $R(u)$ satisfies the Yang-Baxter equation

\begin{equation}
R_{12}(u-v)R_{13}(u)R_{23}(v) = R_{23}(v)R_{13}(u)R_{12}(u-v),
\end{equation}

\noindent where $R_{jk}(u)$ denotes the matrix acting non-trivially
on the $j$-th and the $k$-th spaces and as the identity on the remaining
space.

Next we define the monodromy matrix  $\hat{T}(u)$,

\begin{equation}
\hat{T}(u)= \left( \begin{array}{cc}
 \hat{A}(u) & \hat{B}(u)\\
 \hat{C}(u) & \hat{D}(u)\end{array}\right),\label{monod}
\end{equation}

\noindent such that the Yang-Baxter algebra is satisfied

\begin{equation}
R_{12}(u-v)\hat{T}_{1}(u)\hat{T}_{2}(v)=\hat{T}_{2}(v)\hat{T}_{1}(u)R_{12}(u-v).\label{RTT}
\end{equation}

\noindent To obtain a solution for the two-site Bose-Hubbard model (\ref{ham}) we need to choose a realization for the monodromy matrix $\pi(\hat{T}(u))=\hat{L}(u)$. In this construction, the Lax operator $\hat{L}(u)$  has to satisfy the algebra

\begin{equation}
R_{12}(u-v)\hat{L}_{1}(u)\hat{L}_{2}(v)= \hat{L}_{2}(v)\hat{L}_{1}(u)R_{12}(u-v),
\label{RLL}
\end{equation}
\noindent where we use the standard notation.

Then, defining the transfer matrix, as usual, through

\begin{equation}
\hat{t}(u)= \mbox{Tr} \;\pi(\hat{T}(u)) = \pi(\hat{A}(u) + \hat{D}(u)),
\label{trTu}
\end{equation}
\noindent it follows from (\ref{RTT}) that the transfer matrix commutes for
different values of the spectral parameter.

We are using the well known \cite{GSantos11a} Lax operator, solution of the equation (\ref{RLL}), 
\begin{equation}
\hat{L}_{i}(u)=
\left(\begin{matrix}
u\hat{I} + \eta \hat{N}_i & \hat{a}_i \\
\hat{a}_i^{\dagger} & \eta^{-1}\hat{I}
\end{matrix}\right)\;\;\;\;\;\; i=1,2,
\label{Lax1}
\end{equation}
\noindent for the boson operators $\hat{a}_i^{\dagger}$, $\hat{a}_i$, and $\hat{N}_i$. These operators obey the standard canonical boson commutation rules. 

Using the co-multiplication property of the Lax operators (\ref{Lax1}) we get the following realization for the monodromy matrix,
\begin{equation}
\pi(\hat{T}(u)) = \hat{L}_{1}(u + \omega ) \hat{L}_{2}(u - \omega),
\end{equation}
\noindent whose entries are,
\begin{eqnarray}
\pi(\hat{A}(u)) &=& (u^2-\omega^2) \hat{I} + \eta u \hat{N} \nonumber \\
&+&\eta ^2 \hat{N}_1\hat{N}_2 -\eta\omega
(\hat{N}_1-\hat{N}_2) +\hat{a}^\dagger_2 \hat{a}_1, \label{A} \\
\pi(\hat{B}(u)) &=& (u+\omega+\eta \hat{N}_1)\hat{a}_2+\eta^{-1}\hat{a}_1, \label{B}\\
\pi(\hat{C}(u)) &=&  (u-\omega+\eta \hat{N}_2)\hat{a}^\dagger_1 + \eta^{-1}\hat{a}^{\dagger}_2, \label{C} \\
\pi(\hat{D}(u)) &=& \hat{a}_1^{\dagger}\hat{a}_2+\eta^{-2} \hat{I}. \label{D}
\label{realiz1}
\end{eqnarray}
\noindent Hereafter we will use the same symbol for the operators and its respective realization, so we define $\pi(\hat{O}(u))\equiv \hat{O}(u)$ for any operator in the entries of the monodromy matrix (\ref{monod}).

The parameters of the Hamiltonian (\ref{ham})  are all real numbers, $K$, $\Delta\mu$, ${\cal E}_J \in \mathbb{R}$. The  parameters in the operators (\ref{A},\ref{B},\ref{C},\ref{D}) can be complex numbers, $u$, $\eta$, $\omega \in \mathbb{C}$, but in this case the transfer matrix is not Hermitian. We will only consider the Hermitian case.

We can apply the algebraic Bethe ansatz method, using the  Fock vacuum as the pseudo-vacuum $|0\rangle = |0\rangle_1\otimes|0\rangle_2$, to find the BAE,
\begin{equation}
\eta^2 (v^2_i -\omega^2) =
\prod ^N_{j \neq i}\frac {v_i -v_j - \eta}{v_i - v_j + \eta}, \;\;\;\;\;i,j = 1,\ldots,N.
\label{becbae} 
\end{equation}

\section{Bethe vectors states}

 In the algebraic Bethe ansatz method, the BVS are constructed by the application of the $\hat{C}$-operator to the pseudo-vacuum $|0\rangle$,

\begin{equation}
|\Psi\rangle=\prod_{j=1}^{N}\;\hat{C}(v_{j})\;|0\rangle,
\end{equation}
\noindent where the $\{v_j\}_1^N$ are solutions of the BAE (\ref{becbae}).

  Using the $\hat{D}$-operator (\ref{D}) we can write the $\hat{C}$-operator (\ref{C}) as,

\begin{equation}
\hat{C}(v_{j})=f_j\hat{a}_{1}^{\dagger} + \mathfrak{\hat{\mathcal{D}}},
\end{equation}
\noindent where
\begin{equation}
f_j \equiv v_{j} - \omega,
\end{equation}
\noindent and
\begin{equation}
\mathfrak{\hat{\mathcal{D}}}\equiv  \eta \; \hat{a}_{2}^{\dagger}\hat{D} = \eta \;  \hat{a}_{1}^{\dagger}\hat{N}_{2} + \hat{a}_{2}^{\dagger} \; \eta^{-1}.
\label{DOP1}
\end{equation}

Now we can write the BVS in the product form as
\begin{equation}
|\Psi\rangle = \prod_{j=1}^{N} \;[f_{j} \hat{a}_{1}^{\dagger} + \mathfrak{\hat{\mathcal{D}}}]\;|0\rangle,
\label{wf1}
\end{equation}
\noindent or in the summation form as

\begin{eqnarray}
|\Psi\rangle & = & \sum_{n=0}^{N}\;\mathcal{F}_{n}\;(\hat{a}_{1}^{\dagger})^{N-n}\;\mathfrak{\hat{\mathcal{D}}}^{n}\;|0\rangle,
 \label{wf2}
\end{eqnarray}
\noindent with the identification

\begin{equation}
\mathcal{F}_{0}  =  \prod_{j=1}^{N} f_{j}, ~~~~~ 
\mathcal{F}_{N}  =  \sum_{\substack{j_1,j_2,\ldots,j_N=1\\(j_1>j_2>\ldots>j_N)}}^{N} \frac{\mathcal{F}_{0}}{f_{j_1}f_{j_2} \ldots f_{j_N}}.
\end{equation}

To explicitly write the BVS (\ref{wf1}) or (\ref{wf2}) we need to expand the powers of the  $\mathfrak{\hat{\mathcal{D}}}$-operator (\ref{DOP1}). We thus consider the following $n$-th power binomial expansion for any two operators $\hat{X}$ and $\hat{Y}$, with $n\geq 2$, proved by  induction,

\begin{eqnarray}
(\hat{X}+\hat{Y})^{n} &=& \sum_{j=0}^{n}\sum_{\alpha=1}^{\frac{n!}{j!(n-j)!}}\hat{P}_{\alpha}(\hat{X}^{j}\;\hat{Y}^{n-j}) \nonumber \\
                      &=& \sum_{j=0}^{n}\left(\begin{array}{c}
n\\
j
\end{array}\right)\hat{X}^{n-j}\;\hat{Y}^{j} + \mathfrak{f}([\hat{X},\hat{Y}]),
\end{eqnarray}
\noindent with
\begin{equation}
\left(\begin{array}{c}
n\\
j
\end{array}\right)=\frac{n!}{j!(n-j)!},
\end{equation}
\noindent where the $\hat{P}_{\alpha}$-operator stand for the set of all permutations of the $\hat{X}$ and $\hat{Y}$ operators, not respecting the commutation rule between them, and $\mathfrak{f}([\hat{X},\hat{Y}])$ is a function of that  commutation rule. Below we show $\mathfrak{f}([\hat{X},\hat{Y}])$ for two values of $n$:

\begin{itemize}
\item[I]- For $n=2$:
\begin{equation}
\mathfrak{f}([\hat{X},\hat{Y}]) = [\hat{Y},\hat{X}].
\end{equation}

\item[II] - For $n=3$:

\begin{equation}
\mathfrak{f}([\hat{X},\hat{Y}]) = [\hat{Y},\hat{X}^2] + [\hat{Y}^2,\hat{X}] + [\hat{X}\hat{Y},\hat{X}] +  [\hat{Y},\hat{X}\hat{Y}].
\end{equation}

\end{itemize}

If the operators commute, $\mathfrak{f}([\hat{X},\hat{Y}]) = 0$, we clearly get the standard commutative binomial formula. In this case the $n$-th power of the $\hat{D}$-operator (\ref{D}) is given by
 
\begin{eqnarray}
\hat{D}^n &=& \sum_{j=0}^{n}\left(\begin{array}{c}
n\\
j
\end{array}\right)(\hat{a}_1^{\dagger}\hat{a}_2)^{n-j}\; \eta^{-2j}.
\label{DK}
\end{eqnarray}
\noindent The $n$-th power of the $\hat{D}$-operator (\ref{DK}) is acting in the pseudo-vacuum as

\begin{equation}
\hat{D}^n \; |0\rangle = \eta^{-2n}\;|0\rangle.
\end{equation}

Redefining the  $\mathfrak{\hat{\mathcal{D}}}$-operator (\ref{DOP1}) as 

\begin{equation}
\mathfrak{\hat{\mathcal{D}}}  =  \hat{X}+\hat{Y},
\end{equation}
\noindent with 

\begin{equation}
\hat{X} = \eta\hat{a}_{1}^{\dagger}\hat{N}_{2},\;\;\;\;\;\hat{Y} = \eta^{-1}\hat{a}_{2}^{\dagger},
\label{xy}
\end{equation}
\noindent we get the binomial expansion of the  $\mathfrak{\hat{\mathcal{D}}}$-operator (\ref{DOP1}) for the $n$-th power,
\begin{equation}
\mathfrak{\hat{\mathcal{D}}}^{n} = \sum_{j=0}^{n}\sum_{\alpha=1}^{\frac{n!}{j!(n-j)!}}\;\hat{P}_{\alpha}(\hat{X}^{j}\;\hat{Y}^{n-j}).
\label{ExpD}
\end{equation}

Applying the $n$-th power of the  $\mathfrak{\hat{\mathcal{D}}}$-operator  (\ref{ExpD}) to the pseudo-vacuum we get 

\begin{equation}
\mathfrak{\hat{\mathcal{D}}}^{n} = \sum_{j=0}^n C_{n,j} |j\rangle_{1}\otimes| n - j\rangle_{2},
\label{Dwf}
\end{equation}
\noindent where the coefficients $C_{n,j}$ satisfies the recursion equation

\begin{equation}
C_{n+1,j} = \eta \;\sqrt{j} \; (n + 1 - j)\;C_{n,j-1} + \eta^{-1}\sqrt{n + 1 - j}\; C_{n,j},
\label{RE1}
\end{equation} 
\noindent with the condition $C_{j,j} = 0$.

The solution of the recursion equation (\ref{RE1}) is 

\begin{equation}
C_{n,j} =  \eta^{2j - n} \sqrt{\frac{j!}{(n-j)!}}\sum_{l=0}^{n-j} (-1)^{n+l-j}\; l^n \left(\begin{array}{c}
n - j\\
l
\end{array}\right).
\end{equation}

Finally, using the binomial expansion of the $\mathfrak{\hat{\mathcal{D}}}$-operator (\ref{Dwf}) we can write the BVS (\ref{wf1}) as
\begin{eqnarray}
|\Psi\rangle & = & \mathcal{F}_{0}\;\sqrt{N!}\;|N\rangle_{1}\otimes|0\rangle_{2} \nonumber\\ 
&+& \sum_{n=1}^{N}\sum_{j=0}^{n}\;\mathcal{F}_{n} \; C_{n,j} \; \sqrt{\frac{(N - n + j)!}{j!}}  \; \nonumber \\ 
&\times & |N - n + j\rangle_{1}\otimes| n - j\rangle_{2}.
\label{wf3}
\end{eqnarray}

The scalar product between one on-shell and one generic off-shell BVS  (\ref{wf3}), $|\Psi\rangle$ and $|\tilde{\Psi}\rangle$, with the set of solutions of the BAE (\ref{becbae}) $\{v_j\}_1^N$ and the generic set $\{\tilde{v}_j\}_1^N$ \cite{ragoucy-su3},  is

\begin{eqnarray}
\langle \tilde{\Psi}|\Psi\rangle &=& N! \; \tilde{\mathcal{F}}^{*}_{0}\;\mathcal{F}_{0} \nonumber \\
&+& \sum_{r,n=1}^{N}\sum_{j=0}^{n}\;\tilde{\mathcal{F}}^{*}_{r}\;\mathcal{F}_{n}\;C_{r,r-n+j}\;C_{n,j} \nonumber \\
&\times &\; \frac{(N-n+j)!}{\sqrt{j!(r-n+j)!}}.
\label{prod-scalar}
\end{eqnarray}

From the scalar product (\ref{prod-scalar}) we can write the norm of the BVS (\ref{wf3}),

\begin{eqnarray}
 \langle\Psi|\Psi\rangle &=& N!\;|\mathcal{F}_{0}|^2  + \sum_{r,n = 1}^{N} \sum_{j=0}^{n} \; \mathcal{F}^{*}_{r}\;\mathcal{F}_{n}\;C_{r,r-n+j} \; C_{n,j} \nonumber \\
 &\times & \; \frac{(N-n+j)!}{\sqrt{j!(r-n+j)!}} .
\end{eqnarray}

Because the total number of atoms is a conserved quantity all the BVS (\ref{wf3}) are eigenfunctions of the  $\hat{N}$-operator with the same eigenvalue $N$, and so they are degenerate states for this operator. As $\hat{N}_1$ and $\hat{N}_2$ are not conserved quantities the BVS are not eigenfunction of these operators but we can still write the form factors  of these operators using the BVS (\ref{wf3}). For instance, the non normalized form factors of the imbalance current between the two BEC is written as

\begin{eqnarray}
\langle \tilde{\Psi} | \frac{\hat{N}_1 - \hat{N}_2}{N} | \Psi \rangle  &=&  N! \; \tilde{\mathcal{F}}^{*}_{0}\;\mathcal{F}_{0}  +  \sum_{r,n=1}^{N}\sum_{j=0}^{n} \tilde{\mathcal{F}}^{*}_{r}\;\mathcal{F}_{n} \nonumber \\
& \times & C_{r,r-n+j}\;C_{n,j}\; \left[1-\frac{2(n-j)}{N}\right] \nonumber \\
& \times & \; \frac{(N-n+j)!}{\sqrt{j!(r-n+j)!}}. \nonumber \\  
\end{eqnarray}

\section{Summary} 
 
  We have explicitly written the Bethe vectors states (BVS) by the algebraic Bethe ansatz method using the $gl(2)$-invariant $R$-matrix and an  algebraic relation between the $\hat{C}$ and the $\hat{D}$ operators. We use a binomial expansion of the $n$-th power of the sum of two operators obtained from that algebraic relation to write a recursion equation and solve it.  The binomial expansion of the $n$-th power of the sum of two operators can be written as a commutative binomial expansion plus a function of the commutator of the operators. We calculate the scalar product and the norm of those BVS. The BVS are degenerate eigenfunctions of the total number of particles $\hat{N}$-operator with eigenvalue $N$. As an example of application of the BVS we calculate the form factors for the imbalance current  operator.

\section*{Acknowledgments}
The authors acknowledge Capes/FAPERJ (Coordena\c{c}\~ao de Aperfei\c{c}oamento de Pessoal de N\'{\i}vel Superior/Funda\c{c}\~ao de Amparo \`a Pesquisa do Estado do Rio de Janeiro), CNPq (Conselho Nacional de Desenvolvimento Cient\'{\i}fico e Tecnol\'{o}gico) and the Ewha Womans University by the grant WCU no. R32-2008-000-101300 for the financial support.

\newpage

\end{document}